\documentclass[12pt,linenumbers, preprint]{aastex}

\usepackage{amsmath,longtable,natbib}
\usepackage{float}
\usepackage{hyperref}
\usepackage{adjustbox}
\usepackage{caption}
\usepackage{graphicx,epstopdf}
\DeclareGraphicsExtensions{.ps}
\shorttitle{Wide-band timing of GMRT MSPs}
\shortauthors{Sharma et al.}
\usepackage{subfig}
\begin{document}
\captionsetup[figure]{labelfont={bf},labelformat={simple},labelsep=period,name={Figure}}
\captionsetup[table]{labelfont={bf},labelformat={simple},labelsep=period,name={Table}}
\title{The GMRT High Resolution Southern Sky Survey for pulsars and transients -- V: Localisation of two millisecond pulsars}
\author{
Shyam~S.~Sharma\altaffilmark{1}, 
Jayanta Roy\altaffilmark{1},
Sanjay Kudale\altaffilmark{1}
Bhaswati Bhattacharyya\altaffilmark{1},
Arpit K. Behera\altaffilmark{1},
Shubham Singh\altaffilmark{1}}
\altaffiltext{1}{National Centre for Radio Astrophysics, Tata Institute of Fundamental Research, Pune 411007, India}
\affil{}
\section*{ABSTRACT}
     The timing follow-up of newly discovered millisecond pulsars (MSPs) is hindered by the larger positional uncertainty (a few tens of arc-minutes) associated with the discovery. In this paper, we present the localization of two MSPs, discovered by the GMRT High Resolution Southern Sky (GHRSS) survey, up to arc-second accuracy using a 33~MHz offline coherently dedispersed gated correlator. This gated correlator is an upgraded version of the earlier 16~MHz design. This new development with a factor of two enhancement in the observing bandwidth offers better sensitivity in the image domain, leading to more precise localization. Aided by the precise position, we followed up these two MSPs with sensitive phased-array (PA) beams of upgraded GMRT from 300 to 1460 MHz. More sensitive observations in the PA mode for these two MSPs yield precise ($\sim$ sub-$\mu s$) time-of-arrivals, with DM uncertainties in the range of $10^{-4}-10^{-5}$ $pc\,cm^{-3}$. We also report the profile evolution of the two MSPs over 300$-$1460 MHz. Finally, we discuss the suitability of these MSPs for the pulsar timing array experiments aimed to detect low-frequency gravitational waves.

\section{Introduction}
\label{sec:intro}
Pulsars, the nature’s best clocks, are excellent laboratories to test the laws of physics under extreme magnetic fields, density and gravity. 
Millisecond pulsars (MSPs) are a subclass of pulsars with a few millisecond period ($<$ 30 ms) and remarkable rotational stability, allowing them to be used as probes for understanding pulsar intrinsic properties as well as probing the intervening interstellar medium (ISM). The long-term timing studies of these MSPs can precisely measure the unmodelled variations (called timing noise) in the time-of-arrival (ToA) of the pulses. 

The Pulsar Timing Array (PTA) (e.g., \cite{1979ApJ...234.1100D}) experiments use a set of MSPs with precise timing parameters to detect Gravitational wave (GW) signals using the angular correlation between the residuals of the arrival times of pairs of pulsars (\cite{1983ApJ...265L..39H}).{The GW frequency range to which PTA is sensitive is 1/baseline to 1/cadence \citep{2018RSPTA.37670293S}. Here, the cadence is the regularity of the observations ($\sim$ weekly) and the baseline is the overall timing observation span of pulsars ($\sim$ tens of years). This weekly cadence and years of timing baseline correspond to the frequency range $10^{-6}-10^{-9}$ Hz, commonly known as the nanoHertz range.} The majority of the nHz GW spectrum is assumed to be contributed by an ensemble of merging supermassive black hole binaries, which forms an isotropic stochastic GW spectrum (\cite{2019A&ARv..27....5B}). The number of MSPs used in the experiment has the strongest influence on the detectability of nHz GW background signal in timing data (\cite{2013CQGra..30v4015S}). This necessitates the identification of newly discovered MSPs suitable for inclusion in PTAs. However, such high precision timing follow-up of newly discovered MSPs are hindered by the large positional uncertainties associated with the discoveries from surveys with single dishes (e.g., GBNCC survey \citep{2014ApJ...791...67S}, HTRU survey \citep{2010MNRAS.409..619K}, PALFA survey \citep{2006ApJ...637..446C} etc.) or with incoherent array (IA) beam of interferometric array (e.g., GHRSS survey; \cite{2016ApJ...817..130B}).
 
{The GMRT High Resolution Southern Sky (GHRSS) survey (\cite{2016ApJ...817..130B}, \cite{2019ApJ...881...59B}, and \cite{2022ApJ...934..138S}) is a low-frequency survey for radio pulsars and transients. The survey uses an IA beam of the full GMRT array in band-3 (300$-$500 MHz) to target the sky away from the galactic plane ($|b|>3^\circ$). This survey has so far detected 30 new neutron stars, including three-millisecond pulsars, two weakly recycled pulsars, two rotating radio transients (RRATs), and 23 normal pulsars\footnote{http://www.ncra.tifr.res.in/$\sim$bhaswati/GHRSS.html}}.
The discovery positional error of pulsars in the GHRSS survey can be order of a degree at 400 MHz. However timing follow-up with sensitive phased array (PA) beams of the GMRT using 70\% of the array warrants $\leq 1'$ positional accuracy. The time needed to determine precise position from timing studies of newly discovered pulsars can take more than a year, with observations having regular cadence. Moreover, the covariance between position and pulsar period derivative ($P$) limits the convergence of the timing fit. In case the pulsar is in binary (specially with longer orbital period), the binary model can also have large covariance with position. The effects of such large covariance in timing fit can be minimized with a precise a priori astrometric position accelerating the convergence to an accurate timing model. Such precise arc-sec localisation can be achieved with interferometric imaging where pulsars are effectively seen as point sources. \cite{2013ApJ...765L..45R} reported the arc-sec localization of five MSPs discovered in a \textit{Fermi}-directed survey (\cite{2013ApJ...773L..12B}, \cite{2015ApJ...800L..12R}, \cite{2021ApJ...910..160B}, and \cite{2022ApJ...933..159B}) using gated imaging. Using these precise positions, \cite{2022ApJ...933..159B} demonstrated phase coherent timing solutions for three of those MSPs (J1120$-$3618, J1646$-$2142, and J1828$+$0625) using the legacy GMRT. Along with a GHRSS MSP (J2144$-$5237), these three MSPs were followed by \cite{2022arXiv220104386S} using the upgraded GMRT (uGMRT). \cite{2022arXiv220104386S} reported $\sim$ 3.4 times better timing precision for these MSPs with the uGMRT wide-band receiver observations compared to the legacy GMRT narrow-band observations reported in \cite{2022ApJ...933..159B} and investigated the possibility of using them in the PTA experiment.

Recently, we discovered two new MSPs in the GHRSS survey pointing of J1243$-$47 and J2059$-$48. The two pulsars were identified in 10 minutes of IA beams pointing towards 12h43m00s, $-$47d00$'$00$''$ and 20h49m00s, $-$48d00$'$00$''$. Both pulsars have rotational periods less than 10 ms. The discovery positional error box associated with the two MSPs is of the order of a square degree. The precise positions of the pulsars are required to characterise their properties through follow-up timing experiments using phased array (PA) beam.
 
In this paper, we demonstrate localization of the two GMRT discovered MSPs within arc-second range using recently upgraded 33 MHz offline gating correlator. We explain the observations in Section-\ref{sec:Section-2}. The localisation methodologies based on pulsar-gated imaging and multiple PA beam formation are described in Section-\ref{sec:Section-3}. We show the localisation results from the two techniques in Section-\ref{sec:Section-4}. We report follow-up studies of the localised MSPs using uGMRT band-3 (300$-$500 MHz), band-4 (550$-$750 MHz), and band-5 (1060$-$1460 MHz) to estimate their ToA and dispersion measure (DM) uncertainties at various frequencies while discussing the possibility of including them in the PTA experiments. In Section-\ref{sec:Section-5}, we summarise the paper. 
 
 \section{Observations}
\label{sec:Section-2}

The localisation analysis using gated imaging and/or multiple PA beamformation uses base-band data recorded with the legacy GMRT system (GSB, \cite{2010ExA....28...25R}). The base-band observation details for GMRT MSPs are listed in Table \ref{Observational_details}. For continuum imaging of the GHRSS field J1243$-$47, we recorded 2.6 s visibilities with 4096 channels from the upgraded GMRT (GWB, \cite{2017JAI.....641011R}) in band-3. After localisation we followed up on the two MSPs with sensitive PA beam in GWB band-3, band-4, and band-5. The details of the continuum observation as well as of the follow-up observations are given in Table \ref{Observational_details}.

\begin{table}[H]
    \centering
    \begin{tabular}{|c|c|}
    \hline
    Base-band recording with GSB & Continuum Imaging with GWB\\
    \hline
    \multicolumn{2}{|c|}{(for localisation)}\\
    \hline
    $N_{antennas}$ = 30 & $N_{antennas}$ = 30\\
    Frequency = 306$-$339 MHz (GSB band-3) & Frequency = 300$-$500 MHz (GWB band-3)\\
    $T_{sample}$ = 15 ns & $T_{sample}$ =  2.6 s \\
    Duration = 1 h (J1243$-$47) & Duration = 1 h (J1243$-$47) \\
    ~~~~~~~~~= 45 min (J2059$-$48) & \\ 
    Data size = 3.5 TB (J1243$-$47) & $N_{baselines}$ = 435 \\
        ~~~~~~~~~~~~~~= 2.5 TB (J2059$-$48) & $N_{channels}$ = 4096\\
    \hline
    \end{tabular}
  
    \begin{tabular}{|c|c|c|c|c|c|}
    \hline
    \multicolumn{6}{|c|}{GWB PA beam observations (for timing observations)}\\
    \hline
    Reciever &Mode& Frequency& Usable& $T_{sample}$ & $N_{antennas}$ \\
    Band &  & coverage (MHz) & bandwidth (MHz) & ($\mu s$) & / $N_{channels}$ \\
    \hline
    Band-3 & C & 300$-$500 & 140 & 10.24& 22 / 512 \\
    Band-4 & C & 550$-$750 & 134 & 10.24 & 25 / 512 \\
    Band-5& I & 1060$-$1460& 356 & 81.92 & 27/4096 \\
    \hline
    
    \end{tabular}
    \caption{The table summarises the details of the GSB base-band, GWB visibility, and GWB PA beam observations used in this paper. Only band-3 (GSB and GWB) data sets are used for the localization of the two pulsars, where the timing observations spanning from band-3 to band-5. C and I correspond to coherently and incoherently dedispersed modes respectively. We used 200 MHz bandwidth for our C mode observations since real-time coherent dedispersion mode with an instantaneous bandwidth of 400 MHz is not currently available in the GMRT.{We have a single PA beam observation in each frequency band.} Table \ref{timing_table_b3/4/5} lists the duration of the PA beam observations.}
    \label{Observational_details}
\end{table}

 We used the coherently dedispersed offline gated correlator to generate beams from 33.33 MHz base-band data, as discussed in next section. The base-band data was processed to form IA/PA beams with 1024 channels for J1242$-$4712 (in the field of J1243$-$47) having dispersion measure (DM) of 78.65 $pc\,cm^{-3}$ and 256 channels for J2101$-$4802 (in the field of J2059$-$48) having DM of 25.05 $pc\,cm^{-3}$. We used \texttt{REALFFT+ACCELSEARCH} of \texttt{PRESTO} \citep{2011ascl.soft07017R} to find the best-fit topocentric period and period derivative(s) from the IA beam data sets. In order to localise a MSP using gated imaging, we generated 2 s gated visibilities folded at best-fit topocentric period and period derivative(s). We used \texttt{CASA} software version 5.6.0-60\footnote{https://casa.nrao.edu/casadocs/casa-5.6.0} \citep{2007ASPC..376..127M} to image both the continuum (from GWB) and the gated (from GSB) visibility.
 
 After successful localisation, in follow-up observations, the PA beams in band-3 and band-4 were recorded with online coherent dedispersion, in which the spectral voltages in each frequency subband are corrected for dispersive delays using the nominal value of DM. In band-5 observation we obtained Stokes-I filterbank data sampled at 81.92 $\mu$s with 4096 channels for offline incoherent dedispersion. For J1242$-$4712 and J2101$-$4802, the intra-channel smearing values in band-5 are 32 and 10 $\mu s$, respectively. Both coherent and incoherent dedispersion mode filterbanks were incoherently dedispersed to remove the inter-channel smearing.
 
 We used \texttt{REALFFT+ACCELSEARCH} to find the best-fit period and period derivatives for each PA beam data set. The PA beam data was then folded in \texttt{PRESTO} using the \texttt{PREPFOLD} command at the best-fit rotational parameters. The beam data from band-3 and band-4 were folded into single time-integration, 512 subbands, and 512 profile bins. In band-5, we created a single subintegration, 64 subbands, and 64 bins. We converted \texttt{PRESTO}'s folded \texttt{PFD} to \texttt{PSRCHIVE} \citep{2012AR&T....9..237V} format using the \texttt{PAM} command. The folded data cube was converted to \texttt{FITS} format using \texttt{PSRCHIVE}'s \texttt{PSRCONV} command. The \texttt{FITS} file format is used for deriving ToA and DM uncertainty.

\section{Localisation techniques}
\label{sec:Section-3}
\subsection{Coherently dedispersed gated imaging}
\label{Gating_technique}A continuum image covering an area more than $\sim 1^{\circ}\times 1^{\circ}$ (similar to half-power-beam-width at the lower edge of band-3) in the sky can contain hundreds of point sources. Identification of the pulsed emission is difficult when multiple continuum point sources are present in the same field. Gated imaging is a method that isolates pulsating point source from continuum sources by binning the visibility signals in pulse longitude allowing to separate the ON emission from OFF-pulse region for each period of the pulsar. By subtracting the OFF emission from ON gate, this approach efficiently removes all background/foreground continuum sources from the ON$-$OFF gated image. In a single observation, it can result in arc-second unambiguous localisation of the pulsar, which appears as a single point source in the gated image. \cite{2013ApJ...765L..45R} provides a detailed discussion of this procedure. Here we briefly explain the steps involved in the processing.

(a) Antenna-based coherent dedispersion and offline beamforming: We record Nyquist-sampled base-band data from individual antennas using the GSB. The base-band data for each antenna is then coherently dedispersed to remove the dispersive delays caused by the line-of-sight electron density. The geometric delay between the antennas is corrected during offline correlation and beamforming of coherently dedispersed voltages. Spectral visibility and an IA beam are generated from the correlator and beamformer respectively.  
  
(b) Best-fit topocentric model: The rotational and orbital parameters (for binary systems) are not precisely known for the newly discovered MSPs. We search for the best-fit topocentric period and period derivative(s) using the IA beam. The IA beam is then folded at the derived parameters to locate the ON and OFF bins in the pulse longitudes.

(c) Gating ON and OFF-pulse visbilities: Coherently dedispersed base-band data are used to construct correlated visibility, which are binned and folded individually over the pulse longitude (say $n_{bin}$) using the derived topocentric rotational parameters. The gating correlator produces $n_{bin}$ visibility files corresponding to the bins on the pulse profile. Each bin file contains the complex visibility products as a function of time, frequency, and baselines.
 
(d) Imaging: We generate the optimal number of bins along pulse longitude so that ON-pulse emission is contained inside a single bin or in a few adjacent bins ($n_{ON}$). The OFF-pulse emission is contained within the remaining $n_{bin}-n_{ON}$ bins. The OFF-pulse visibility file(s) are chosen from a set of $n_{bin}-n_{ON}$ files having no contributions from ON-pulse emission (even in case of wide profiles or profile with inter-pulse). In case of multiple ON and OFF bins, the corresponding visibility files are added to form a ON and OFF bin visibility data set. The OFF-pulse visibilities are then subtracted from the ON-pulse visibilities. Following this subtraction, all continuum sources will be eliminated from the ON-OFF visibility file, leaving just pulsar. The gated visibilities of the pulsar field are flagged for bad time and frequency chunks using \texttt{flagcal} \citep{2012ExA....33..157P}. \texttt{flagcal} is a software that flags and calibrates GMRT UV data. The same software is used to calibrate the target field using the data of an external calibrator field observed before the target. 
The position of the pulsar is determined by imaging ON, OFF, and ON$-$OFF gated visibility data (after calibration) using \texttt{CASA}.  \texttt{CASA} is used for deconvolution and self-calibration.
 
\subsubsection{Limitations of the technique}
  
The gated correlator uses narrow-band data with a bandwidth of 16.67 MHz. The availability of lower observational bandwidth for gating makes it difficult to localise fainter MSPs discovered from GHRSS survey over 200 MHz band-width. For low signal-to-noise (S/N) in the image domain, the noise fluctuations in the ON and OFF images might cause false  detection in the ON$-$OFF image, which can imitate pulsars. Also, due to the gating of ON and OFF-pulses on a restricted number of bins ($\sim$ 20), we lose sensitivity for narrow duty-cycle pulsars. 
Since the gating correlator simultaneously writes $n_{bin}$ visibility files to a disk each at a rate of $>$ 2 MB/s, the number of bins is limited by the available disk writing speed.{The time required to generate gated visibilities from 1 hour of baseband data is of few days for MSPs, depending on pulse period and the number of phase bins.} Moreover, pulsars with $\sim$ mJy (or less) flux density and located near the beam's edge are sometime difficult to spot on the ON image.

\subsubsection{Upgrade in bandwidth of correlator}
\label{sec:upgrade}
We implemented a base-band recording capability with 33.33 MHz band-width (twice of the previously supported band-width) with 2 bits per Nyquist sample. The corresponding upgrade was made to the offline correlator, resulting in improved sensitivity for imaging fainter pulsars.

\begin{figure}[H]
    \centering
    \includegraphics[width=0.8\linewidth, keepaspectratio]{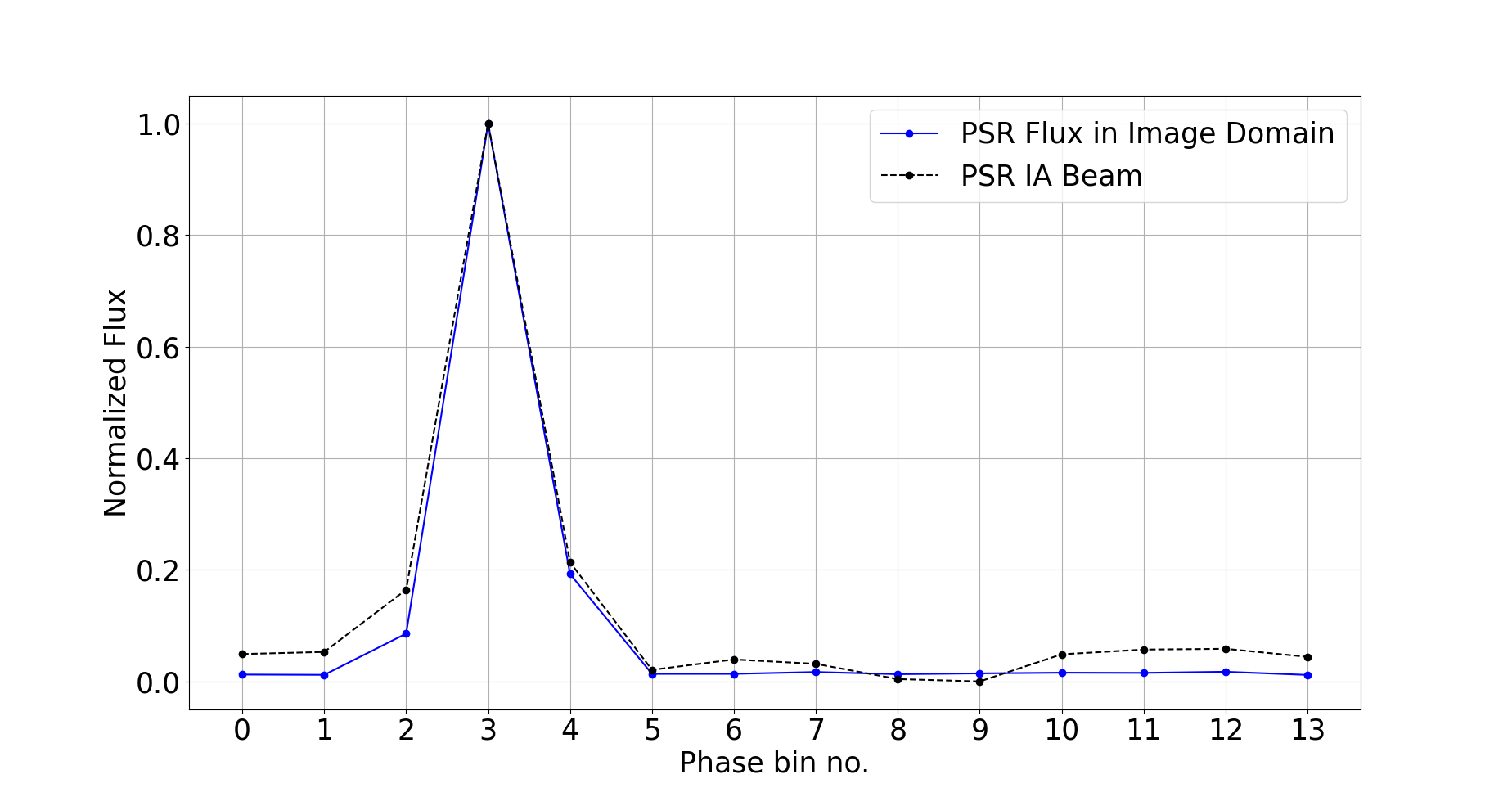}
    \caption{Figure showing B1929$+$10 normalised flux density curve (in blue) derived from gated imaging of 33.33 MHz base-band data. The pulse phase was divided into 14 equal phase bins. We imaged the gated visibility data for each of the 14 bins.{The imaging flux measurements are divided by the maximum detected flux value.} The black curve represents detection in a time-domain IA beam data folded with \texttt{PRESTO}.{We subtracted the minimum value from each point in the IA beam's curve before normalising to 1.}}
    \label{B1929_light-curve}
\end{figure}

In order to validate this newly developed mode, we observed pulsar B1929$+$10 with 33.33 MHz gating correlator. We formed 14 gated visibility files, corresponding to 14 bins of the pulse longitudes, which were imaged for the test pulsar B1929$+$10. Each visibility file yields an estimate of the pulsar's flux density at its location. {We compared the flux density of test pulsar from each visibility file with the pulse profile from simultaneously generated IA beam to validate the functionality of binning and folding during the gating process.} Figure \ref{B1929_light-curve} depicts the normalised flux density at the pulsar location in individual gated images from the upgraded gating correlator. In the same plot, the pulsar IA beam profile produced from a simultaneous beamforming operation over the same 33.33 MHz bandwidth is over-plotted. The similarity of the two curves validates the proper functioning of the 33.33 MHz offline gating correlator without any contamination of OFF-pulsed emission by the ON-pulse energy.

\subsection{Multiple PA beam formation}

Using the GWB band-3 visibility data over 300$-$500 MHz, we produce a continuum image more than $1^{\circ}\times 1^{\circ}$ in size. We used an imaging pipeline comprised of \texttt{flagcal}, \texttt{PyBDSF}, and \texttt{CASA} to analyse uGMRT wideband visibility data (Kudale $\&$ Chengalur, in preparation). The image's phase centre is maintained at the GHRSS discovery pointing. The point sources are extracted from the continuum image using \texttt{PyBDSF} software \citep{2015ascl.soft02007M}, which is a software package for extracting sources from radio images. This results in a list of a few hundred of point sources and their locations. 
\texttt{PyBDSF} decomposes the image into set of gaussians, shapelets, or wavelets. It computes the background rms (root-mean-square), mean images, and locates the emission zone over the user-specified threshold. When Gaussian fitting is applied to these locations, the gaussians are grouped into sources.

We removed the entries for bright (considering the expected flux density of the pulsar obtained from IA beam detection) and extended sources, resulting in a reduced number of sources in the list. Base-band data of 33.33 MHz bandwidth for the same field is recorded at the Nyquist rate. With an offline  correlator, we create multiple PA beams (\cite{2012MNRAS.427L..90R}) using 70$\%$ GMRT array (having a beam size of a few tens of arc-sec), where each beam is directed at individual point sources. One of the point sources would exhibit pulsed detection at the highest significance in its corresponding PA beam. The detection is significantly more pronounced in PA beam with the pulsar in it compared to the IA beam. That point source is then identified as a pulsar and the positional accuracy is derived from the continuum image.{The computation time for this method is comparable to that of the gated imaging technique. Furthermore, gated imaging often includes noise fluctuation alongside the pulsar in the ON-OFF image, making it challenging to pinpoint location of weak pulsars (especially for the field with strong sources creating artefacts). For weak pulsars, multiple PA beam formation is more robust than gated imaging in determining the unique position of the pulsar provided the pulsar is one of the detectable point sources in the continuum image.}

\subsubsection{Limitations of the technique}

 The success of this technique depends upon the detectability of pulsar in the GWB continuum image. Faint sources in the GWB image are often mixed with artifacts in presence of nearby strong sources and may not be listed in the extracted point sources catalog. In addition, creating multiple PA beams from base-band data necessitates a significant amount of processing time. {For example, around a week of processing time for beamforming and folding is required to locate a pulsar from a few hundred of point sources.} Moreover, artifacts near strong sources contribute to the list of point sources, resulting in more computation time.
 
 
\section{Localisation and timing of newly discovered MSPs}
\label{sec:Section-4}
\subsection{Localisation}
        
\begin{figure}[H]
  \centering
      \includegraphics[width=0.49\linewidth, keepaspectratio]{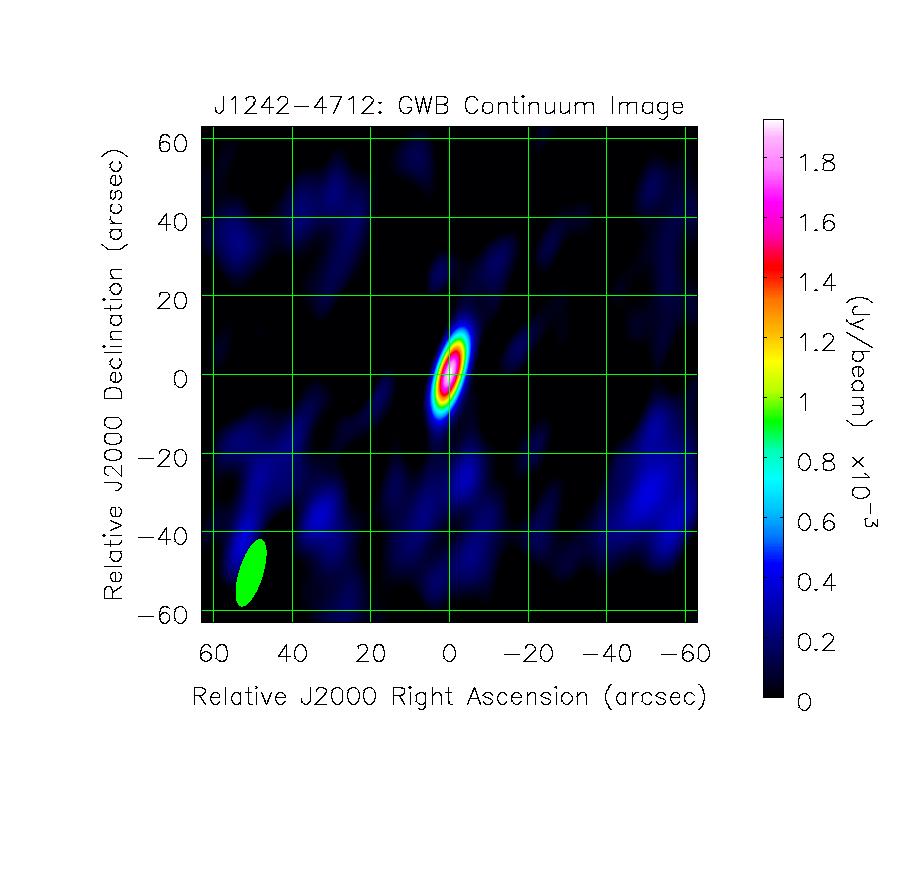}
      \includegraphics[width=0.49\linewidth, keepaspectratio]{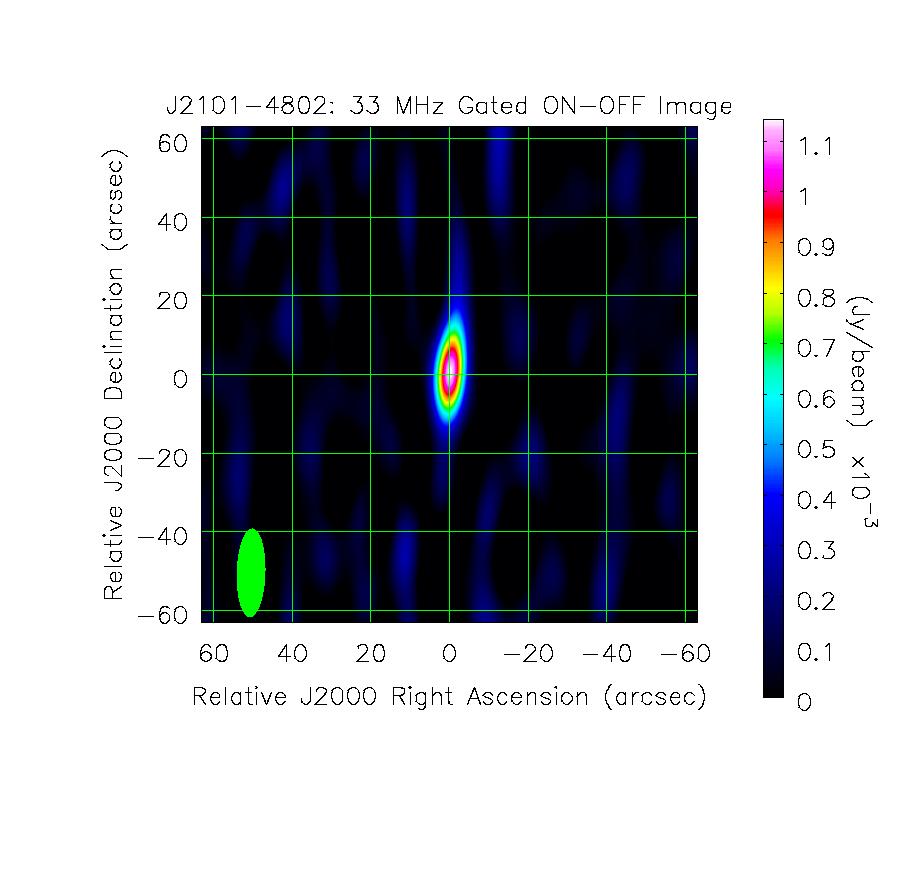}
      \caption{Left: A continuum image of the MSP J1242$-$4712 created using GWB band-3 visibility observations. The pulsar in the image has a S/N of $\sim$ 20, and the image's rms flux value is 0.10 mJy. Right: An ON$-$OFF gated image of the MSP J2101$-$4802 created with the GSB 33.33 MHz base-band data. The image's rms flux is 0.08 mJy, and the pulsar's S/N is 14. We chose two bins out of 19 as ON-pulses and two as OFF-pulses. Two ON-pulses and two OFF-pulses' visibility are added separately. The added OFF-pulse visibilities are subtracted from the added ON-pulse visibilities. The final ON$-$OFF gated image is shown here.{Note that the source location is set as the origin for both images, and the relative axis is plotted. Table \ref{basic_parameter_table} shows the origin's position}.}
      \label{localisation_images}
  \end{figure}
\begin{figure}[H]
\centering
    \subfloat[J1242$-$4712, IA beam\label{GSBa}]{{\includegraphics[width=0.25\linewidth, keepaspectratio]{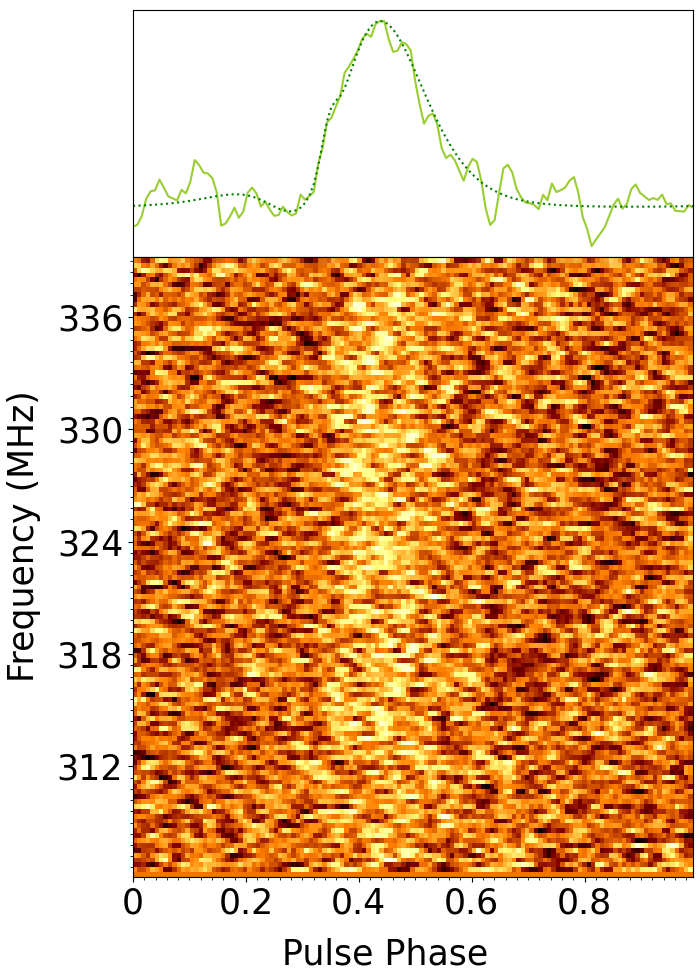} }}
    ~~~~~
    \subfloat[J2101$-$4802, IA beam\label{GSBb}]{{\includegraphics[width=0.25\linewidth, keepaspectratio]{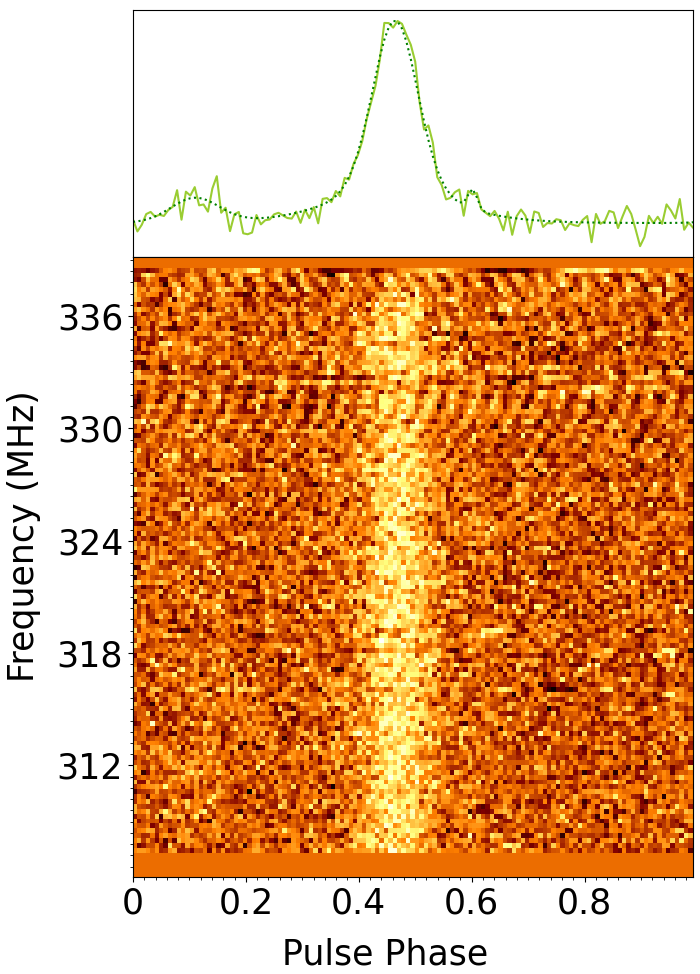} }}\\
    
    \subfloat[J1242$-$4712, PA beam\label{GSBc}]{{\includegraphics[width=0.25\linewidth, keepaspectratio]{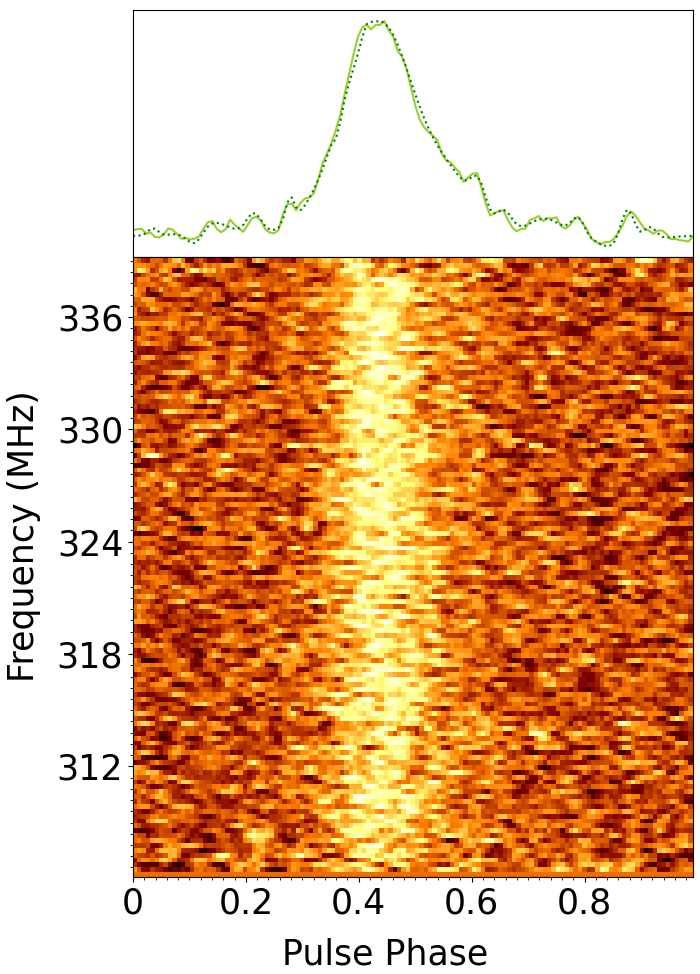} }}
     ~~~~~
    \subfloat[J2101$-$4802, PA beam \label{GSBd}]{{\includegraphics[width=0.25\linewidth, keepaspectratio]{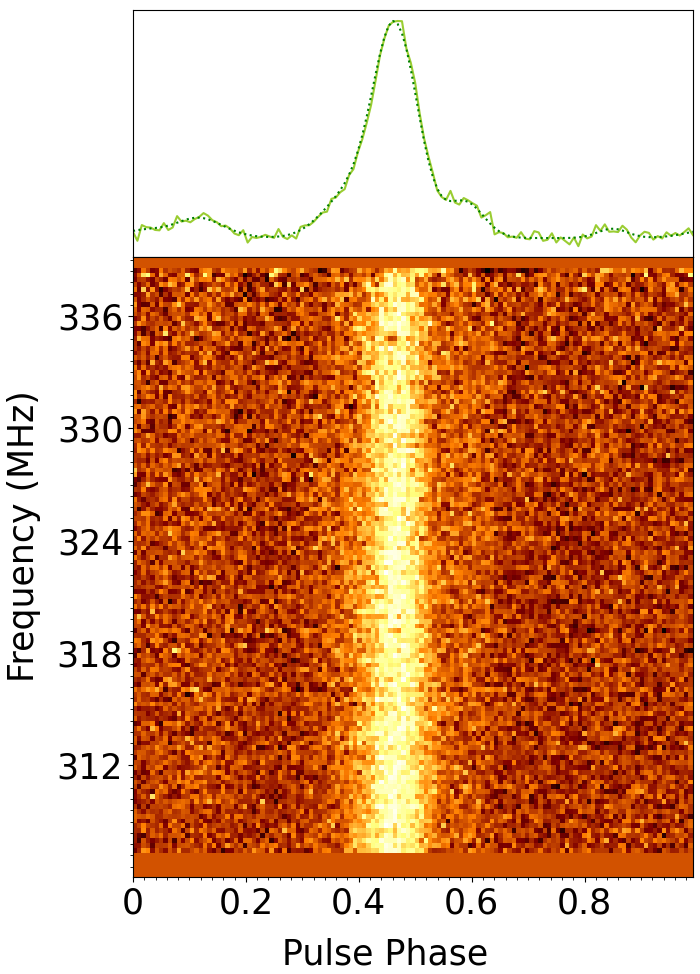} }}
        \caption{Figures showing the {incoherently dedispersed} IA band PA beams at the positions of J1242$-$4712 and J2101$-$4802. The beams were generated from the 33.33 MHz base-band observations using an offline correlator. {We created 1024 and 256 channels in IA/PA beams for J1242$-$4712 and J2101$-$4802, respectively.} The S/N of IA and PA beams for J1242$-$4712 are 36 and 139, respectively. We get 59 and 145 S/N detection in GSB IA and PA beams for J2101$-$4802, respectively.}
    \label{GMRT_pulsar_profiles_GSB}
\end{figure}

 The continuum visibility collected for the field of J1243$-$47 is flagged and calibrated (using 1154$-$350 calibrator) with \texttt{flagcal}. We performed three self-calibration cycles, the first two using only phase and the third using amplitude and phase calibration mode. The \texttt{Pybdsm} based source extraction routine reports 245 point sources above the 5 $\sigma$ level with $\sigma=$ 0.1 mJy within a $1.3^{\circ}\times1.3^{\circ}$ GWB continuum image of J1243$-$47 field. The MSP J1242$-$4712 was found among 245 sources by creating multiple PA beams using base-band data. It was located 18$'$ away from the phase centre of the GWB continuum image, which was the GHRSS survey pointing location. The pulsating point source, identified in the GWB continuum image, is shown on the left panel of Figure \ref{localisation_images}. Figures \ref{GSBa} and \ref{GSBc} show the IA and PA beams generated using base-band data at the location of this point source, where PA beam pointed on the pulsar gives $\sim 3.9$-times more S/N than the IA beam.

We followed a similar strategy to locate J2101$-$4802. The GWB continuum image of this field, with size of $2^{\circ}\times2^{\circ}$, generated after calibrating (using 2225$-$049 calibrator) and flagging, contained 188 point sources. However, none of the PA beams formed using 70$\%$ array gave expected detection significance compared to the IA beam.{One of the PA beams showed very faint detection (even weaker than the IA beam) and provides a possible location of the pulsar with few arc-minutes error.} We used the gating imaging technique to generate visibilities in 19 bins over the pulse longitude{taking the possible location as new phase centre.} The ON and OFF bins were identified from the simultaneously formed IA beam profile. Imaging the ON$-$OFF visibility bin we found the pulsar 51$'$ away from the{original} phase centre of the GWB image {(GHRSS discovery location)} and the nearest point source (in GWB image) where we get the faint detection in PA beam was at a distance of 1.4$'$ away from the location of the pulsar. The distance of 51' from phase centre is outside the full-width-half-maximum (FWHM) of the primary beam in GMRT band-3 (75$'$ at 400 MHz). The GWB continuum image of this field contained no point sources above 5 $\sigma$ (where $\sigma=$ 0.1 mJy), at the gated location. 
Thus a pulsar with flux density of $\sim 1 $ mJy at 400 MHz may not be detected in the continuum image if it is outside the FWHM of primary beam of GMRT. The gated image of the pulsar is shown on the right side of Figure \ref{localisation_images}. Figures \ref{GSBb} and \ref{GSBd} show the IA and PA beams generated using base-band data at the gated location, where detection significance in PA beam is $\sim 2.5$-times more than the IA beam. The J2000 positions, as well as other parameters for the two MSPs, are listed in Table \ref{basic_parameter_table}.

  \begin{table}[H]
    \centering
    \begin{adjustbox}{width=\columnwidth,center}
    \begin{tabular}{|c|c|c|c c c|c|c|}
    \hline
         PSR & Type & P & Frequency & DM & Flux & J2000 Position & Offset from\\
         
         & & $(ms)$  &Band & $(pc\,cm^{-3})$ & $(mJy)$ & ($hh:mm:ss.s$ & survey pointing\\
         & &  & & & &  $dd:mm:ss.s$)&  centre ($arc-min$)\\
         \hline
         J1242$-$4712 & binary & 5.31 & Band-3 &{78.6472 (6)} & 2.00 (8) & 12:42:12.80 (1) & 18\\
          & & & Band-4 &{78.643 (3)} &{1.22 (7)}  & -47:12:17.3 (3) & \\
          & & & Band-5 &{78.71 (8)} &{0.31 (6)} & & \\
         
         J2101$-$4802 & binary & 9.48  & Band-3 &{25.055 (1)} &{10.2 (6)} &{21:01:55.115 (5)} & 51\\
         
          & & & Band-4 &{25.095 (5)} &{4.0 (1)} &{-48:02:02.9 (5)} & \\
          & & & Band-5 &{25.0 (1)} &{1.23 (9)}  & & \\
         \hline
    \end{tabular}
    \end{adjustbox}
    
    \caption{The table shows the basic parameters of the two MSPs.{Using \texttt{PSRCHIVE}'s \texttt{PDMP} utility, we determined the absolute DM values for the two MSPs in different frequency bands.} The {period averaged} flux densities, RA, and DEC coordinates in J2000 for the two MSPs are derived from their GWB continuum images. The last column shows the distance between the discovery position (GHRSS pointing) and the locations are given in the fifth column.}
    \label{basic_parameter_table}
    
\end{table}

\subsection{Follow-up Investigations}

\subsubsection{Integrated Profiles}
\subsubsubsection{J1242$-$4712}

\begin{figure}[H]
    \centering
    \subfloat[J1242$-$4712, Band-3\label{P1a}]{{\includegraphics[width=0.25\linewidth,keepaspectratio]{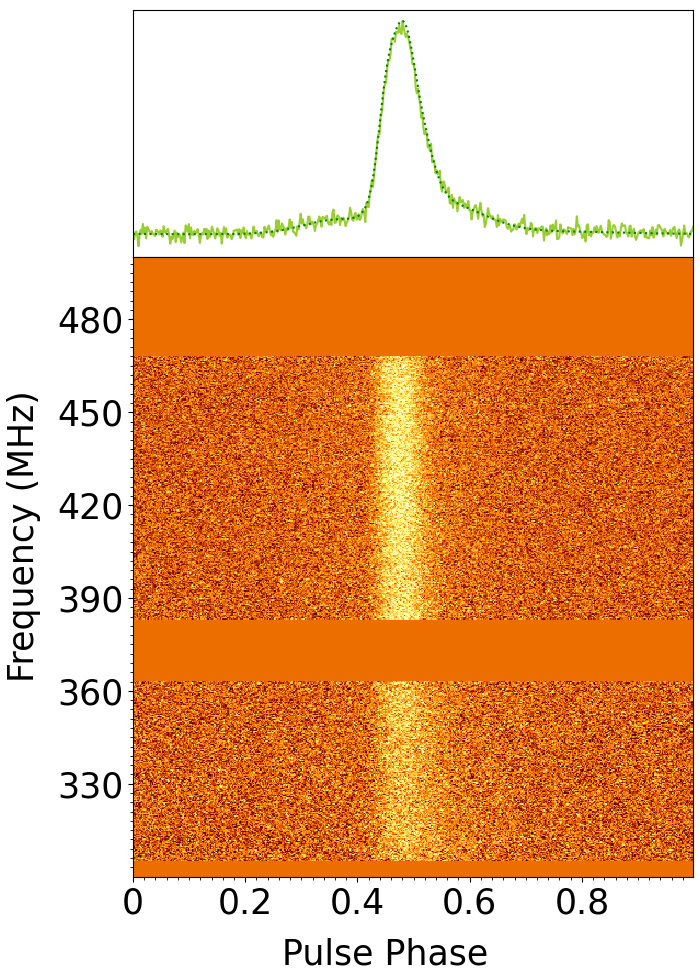} }}
    ~~~~~
    \subfloat[J2101$-$4802, Band-3\label{P2a}]{{\includegraphics[width=0.25\linewidth,keepaspectratio]{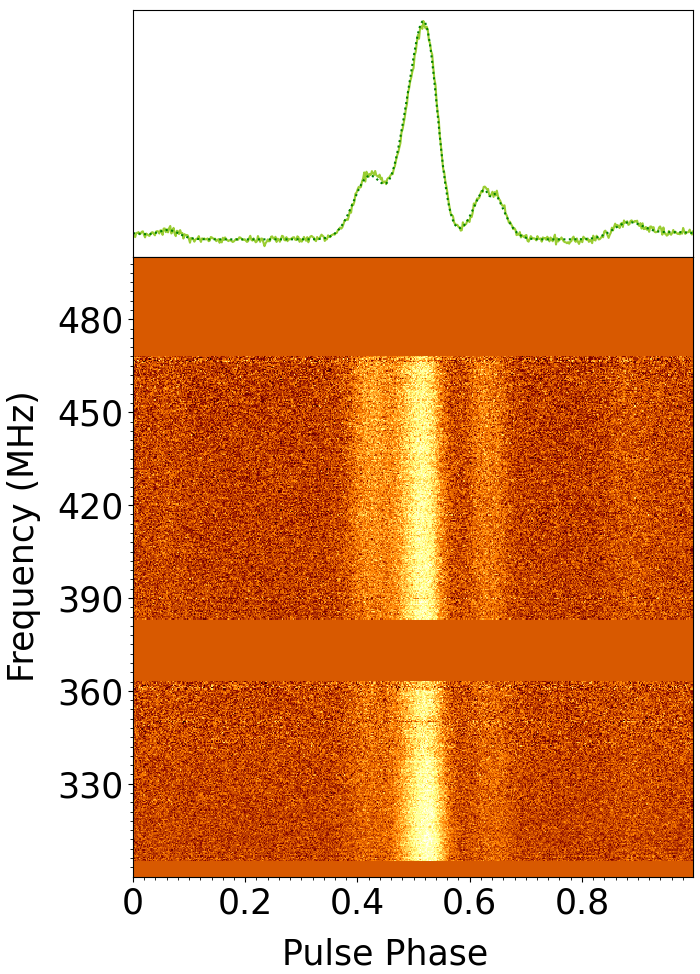} }}\\
  
    \subfloat[J1242$-$4712, Band-4\label{P1b}]{{\includegraphics[width=0.25\linewidth,keepaspectratio]{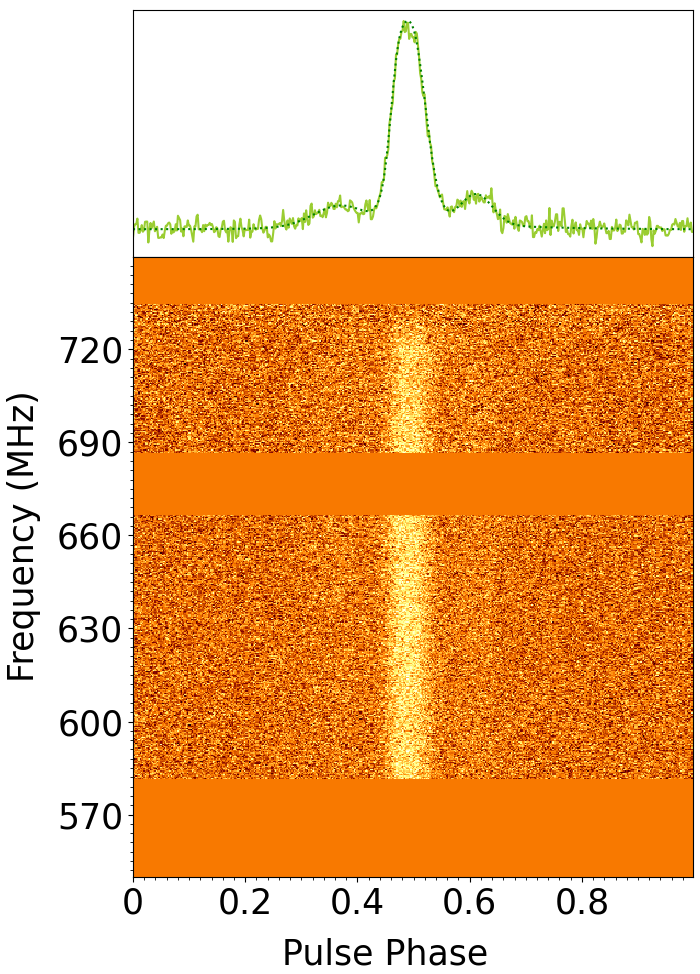} }}
    ~~~~~
    \subfloat[J2101$-$4802, Band-4\label{P2b}]{{\includegraphics[width=0.25\linewidth,keepaspectratio]{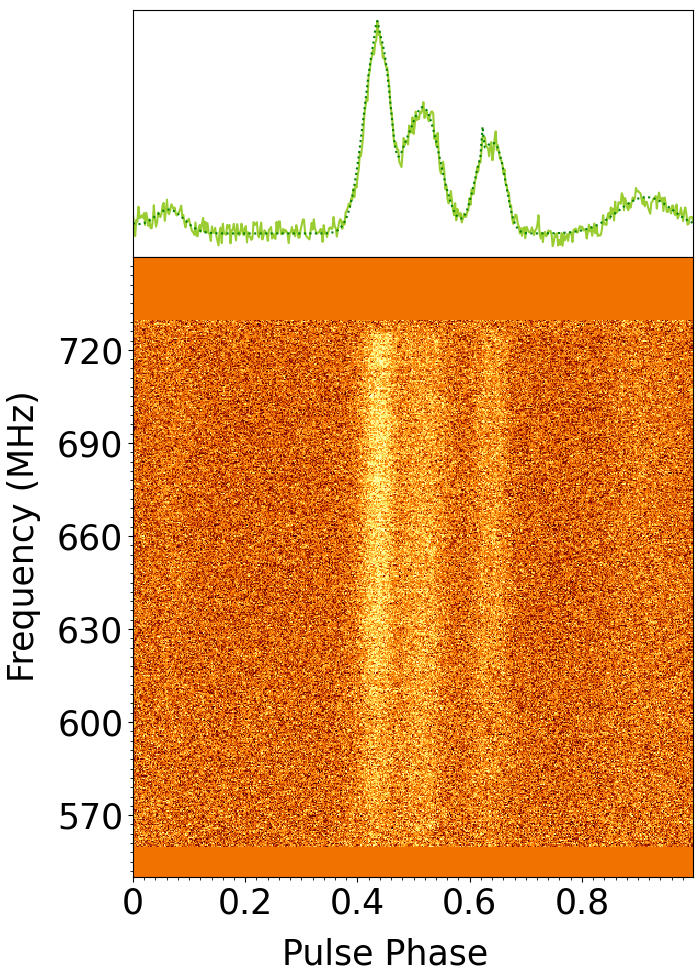} }}\\
     \subfloat[J1242$-$4712, Band-5\label{P1c}]{{\includegraphics[width=0.25\linewidth,keepaspectratio]{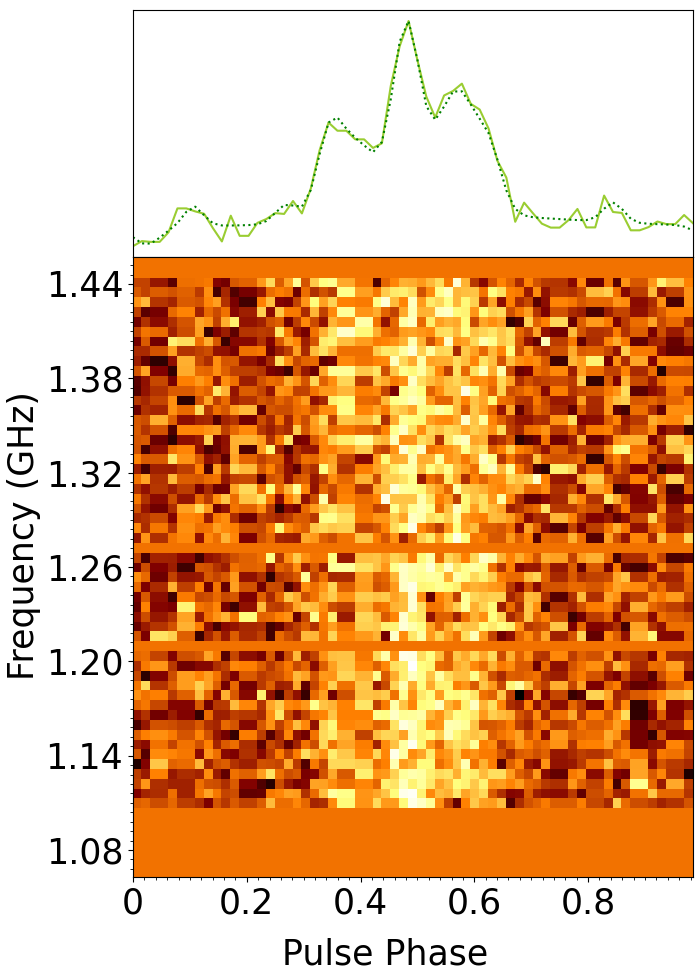} }}
    ~~~~~
    \subfloat[J2101$-$4802, Band-5\label{P2c}]{{\includegraphics[width=0.25\linewidth,keepaspectratio]{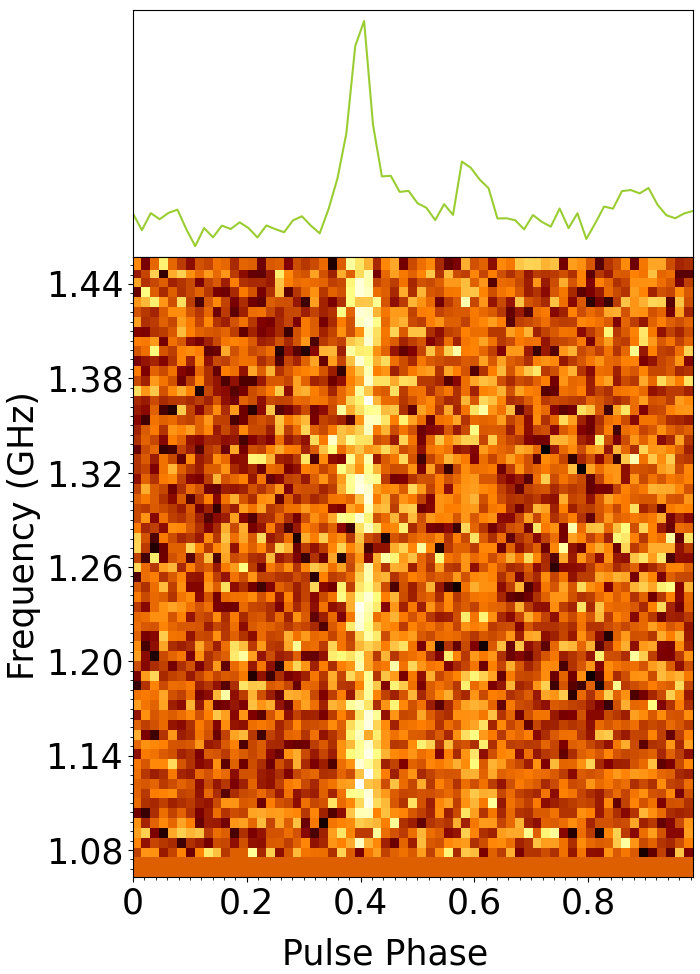} }}
        \caption{Figure showing the integrated profiles and frequency-phase plots of the two MSPs in the GWB bands-3/4/5.{In all frequency bands, a single epoch folded profile is shown.}}
    \label{P1_and_P2_light_curve}
\end{figure}

J1242$-$4712 has a single component profile in band-3 (Figure \ref{P1a}), with W10 and W50 (width at one-tenth and half of peak intensity) of 1010 $\pm$ 10.38 $\mu s$ and 432 $\pm$ 10.38 $\mu s$, respectively. It has three profile components in band-4 (Figure \ref{P1b}). The W50 for the middle (strongest) component is 356 $ \pm$ 10.38 $\mu s$.
This pulsar's band-5 profile (Figure \ref{P1c}) also has three central components, with the middle one being the strongest. The W50 value for the middle component is 456 $ \pm$ 83.02 $\mu s$.

The number of profile components increased from one to three between band-3 and band-4. From band-4 to band-5, the ratio of the third component (from left to right in Figures \ref{P1b} and \ref{P1c}) to the first component changes from 1.24 to 1.33, and the ratio of the second component to the third component changes from 6.16 to 1.53. This pulsar's profile clearly shows significant evolution from band-3 to band-4 to band-5.

\subsubsubsection{J2101$-$4802}

J2101$-$4802 has three strong components near its centre and two weak components at the beginning and end of the profile in band-3 (Figure \ref{P2a}). The W10, which corresponds to the first and second central components, is 1690 $\pm$ 18.52 $\mu s$. The W50 value for the second central component (the strongest) is 629 $\pm$ 18.52 $\mu s$. The number of profile components in band-4 is the same as in band-3 (Figure \ref{P2b}). The W50 for the first central component (the strongest) in band-4 is 482 $\pm$ 18.52 $\mu s$. The profile in band-5 has three central components, the middle of which is almost non-existent (Figure \ref{P2c}). The W50 for the first (the strongest) component is 448 $\pm$ 148.13 $\mu s$.

The sizes of the three central components vary across frequency bands. The ratio of the first to third central component (from left to right in Figures \ref{P2a}, \ref{P2b}, \ref{P2c}) varies from 1.30 to 2.37 to 2.41 from band-3 to band-4 to band-5, respectively. From band-3 to band-4, the ratio of the second to first central component decreases from 3.54 to 0.62, and the second component almost disappears in band-5. This MSP also shows significant profile evolution from one band to the other.

\subsubsection{ToA and DM precision}

 We used $\textit{PulsePortraiture}$\footnote{\label{pulse_portraiture}\url{https://github.com/pennucci/PulsePortraiture}} software (\cite{2016ascl.soft06013P}; \cite{2014ApJ...790...93P}; \cite{2019ApJ...871...34P}) for the timing analysis of the two pulsars, which can estimate ToA and DM from each epoch's observation simultaneously. The wide-band timing technique is illustrated by {\cite{2014ApJ...790...93P} and \cite{2014MNRAS.443.3752L}}, extracts ToAs and DMs using a 2-dimensional (2-D) template that is a function of rotational phase bin and frequency. This 2-D template is created in $\textit{PulsePortraiture}$ using principal-component-decomposition (\cite{2019ApJ...871...34P}). To extract the wide-band ToA and DM from the observational data, we used the same strategy described in \cite{2022arXiv220104386S}. 
 
 In band-3, we have used{only a single epoch} of observation for both MSPs. The S/N is{220} for J1242$-$4712 and{470} for J2101$-$4802. The time-averaged-frequency resolved FITS data is subjected to a principal-component-analysis (PCA). PCA returns{one eigenvector with mean profile for both of the MSPs}.  A linear combination of the smoothed eigenvector and the mean profile (as described in equation 10 of \cite{2019ApJ...871...34P}) is used to create the frequency dependent template. This template is used to extract a single ToA and DM from the epoch FITS (using equations 7 and 8 of \cite{2014ApJ...790...93P}). Table \ref{timing_table_b3/4/5} shows the ToA and DM uncertainty of the observations in band-3.
 
 In band-4 and band-5, a single FITS was available for both MSPs. The FITS file is subjected to PCA, which produces a mean profile with no eigenvectors. The smoothed mean profiles are used to extract the ToA and DM from band-4 and band-5 FITS. Note that different frequency band observations are analysed using their corresponding templates.

 \begin{table}[H]
    \centering
    \begin{tabular}{|c|c|c|c|c|c|}
    \hline
    PSR & Frequency & Duration & S/N & $\sigma_{ToA}$ &  $\sigma_{DM}$\\
    & Band & ($minutes$) & & ($\mu s$) &  ({$10^{-4}\,pc~cm^{-3}$})  \\
    \hline
    \hline
   & Band-3 & 45 & 203 & 0.744 &  0.90 \\
   J1242$-$4712 & Band-4 & 47 & 175 & 0.444 &  3.64 \\
   & Band-5 & 82 & 43 & 4.366 & 1.10$\times10^{2}$ \\
    \hline
    \hline
   & Band-3 &{45} &{470} &{0.539} &  {0.63} \\
   J2101$-$4802 & Band-4 &{60} &{181} &{1.348} &  {1.18$\times10^{1}$} \\
   & Band-5 &{60} &{37} &{5.417} & {1.14$\times10^{2}$} \\
    \hline
    \end{tabular}
    \caption{The table shows the observation duration, S/N of the GWB PA beam, ToA, and DM uncertainty for the two MSPs in band-3/4/5.{ToAs are calculated from a single epoch FITS in band-3/4/5.}}
    \label{timing_table_b3/4/5}
\end{table}

We obtained less than 1 $\mu s$ ToA uncertainty for both MSPs in band-3 observations. The DM precision for the two MSPs in this band is of the order of $10^{-5} \,pc~cm^{-3}$.
The ToA precision for J1242$-$4712 increases from 0.744 to 0.444 $\mu s$, while the DM precision decreases by four times from band-3 to band-4. For J2101$-$4802, the ToA precision decreases from{0.539} to{1.348} $\mu s$, and the DM precision shifts by more than one order (\textbf{$6.3\times10^{-5}$} to $1.18\times10^{-3}\,pc~cm^{-3}$) from band-3 to band-4.
In band-5, J1242$-$4712 and J2101$-$4802 have ToA uncertainty of 4.366 and{5.417} $\mu s$, respectively, with DM uncertainty of the order of $10^{-2}\,pc~cm^{-3}$. For J1242$-$4712, we found that the DM values in band-3 and band-4{(Table \ref{basic_parameter_table})} are similar within $\pm{2}\sigma_{DM}$ of band-4.{The DM values from band-4 and band-5 are similar within $\pm 1\sigma_{DM}$ of band-5. For J2101$-$4802, we find $\pm{8}\sigma_{DM}$ change in DM value from band-3 to band-4 in respect to the DM uncertainty of band-4. The DM values from band-4 and band-5 are similar within $\pm 1\sigma_{DM}$ of band-5}. Note that the{absolute} DM values described here are derived from single epoch observations (using{\texttt{PDMP} command of \texttt{PSRCHIVE}}) in different frequency bands of the uGMRT.

\subsubsection{Comparison of ToA and DM precision with other PTA MSPs}

\begin{figure}[H]
    \centering
    \includegraphics[width=0.49\linewidth, keepaspectratio]{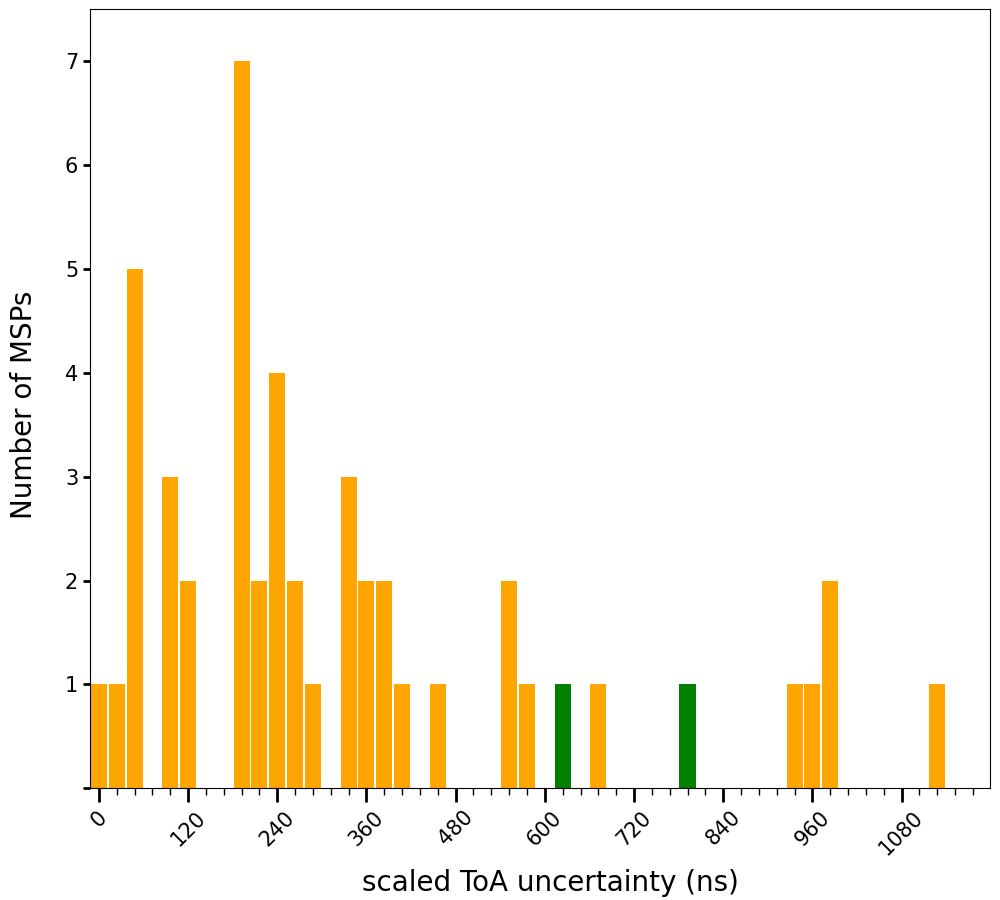}
    \includegraphics[width=0.49\linewidth, keepaspectratio]{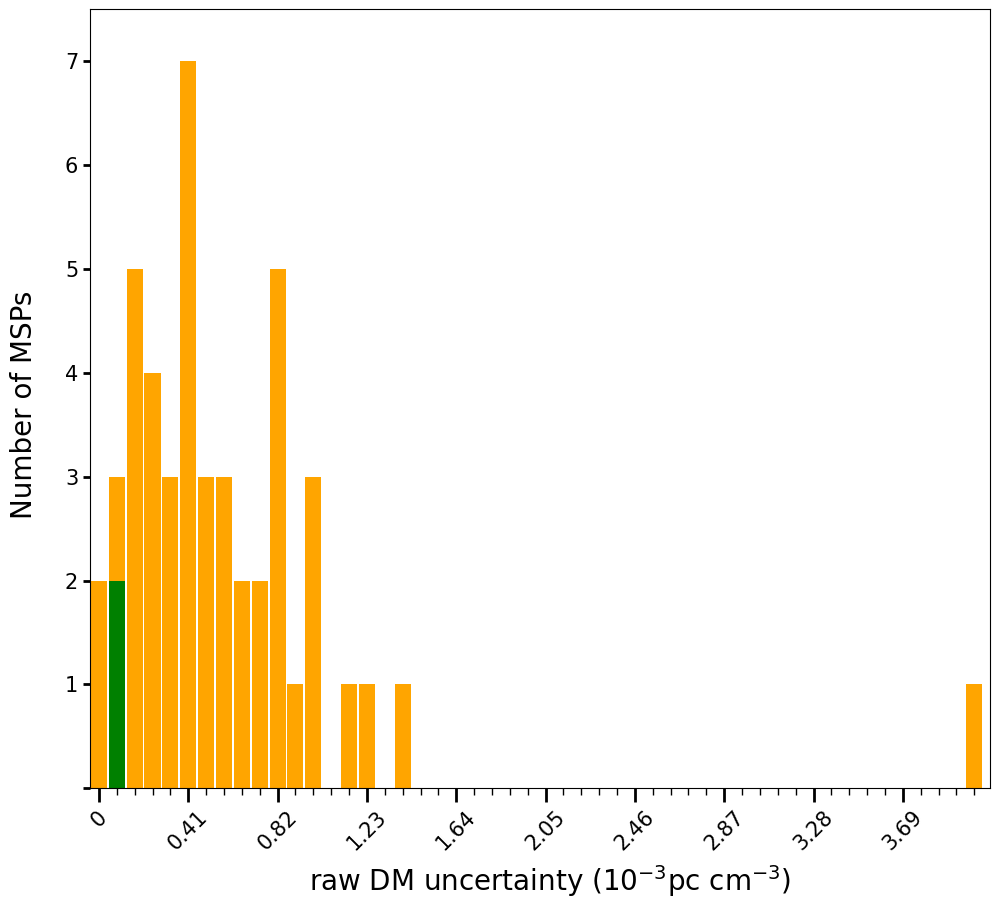}
    \caption{{Figure showing the ToA (left) and DM (right) uncertainty histogram for 47 MSPs (in orange) used in the NANOGrav PTA. We used the median scaled ToA uncertainty and median raw DM uncertainties of 47 PTA pulsars for this plot, which are presented in Table 4 and Figure 2 of \cite{2021ApJS..252....5A}. We chose a single median ToA and DM uncertainty for each MSP based on the highest precision achieved among various telescope receivers. We have scaled the minimum ToA uncertainties for J1242$-$4712 and J2101$-$4802 given in Table \ref{timing_table_b3/4/5} by $\sqrt(\frac{\textbf{\emph{usable}} \,bandwidth}{100 MHz}\times\frac{duration}{30 min})$. The scaled uncertainties for the two MSPs are over-plotted in green in the left panel histogram. For right panel histogram, we have over-plotted the minimum DM uncertainties (raw) from Table \ref{timing_table_b3/4/5} for J1242$-$4712 and J2101$-$4802 (in green).}}
    \label{ToA_uncertainty_histogram_49_MSPs}
\end{figure}

The North American Nanohertz Observatory for Gravitational Waves (NANOGrav; \cite{2009arXiv0909.1058J}) aims to detect stochastic GW background using a set of 47 MSPs. Using $\textit{PulsePortraiture}$ software, \cite{2021ApJS..252....5A} reported the wide-band timing of all of these MSPs observed with the Arecibo and GBT over a frequency range of 315$-$2380 MHz. The median scaled ToA uncertainty and median raw DM uncertainty of each of these MSPs in various frequency bands are given in Table 4 and Figure 2 of \cite{2021ApJS..252....5A}. We chose the most precise median ToA and DM uncertainty among these different frequency bands for each MSP and plotted the histograms as seen in Figure \ref{ToA_uncertainty_histogram_49_MSPs}. In the same histograms, we have added the best ToA and DM precision obtained for J1242$-$4712 and J2101$-$4802. 
{This shows that the scaled ToA precision 
for both MSPs are well within the NANOGrav samples and certainly they have very high DM precision (within top 10\% of NANOGrav measurements). This illustrates the prospect of both these newly discovered pulsars for inclusion in the PTAs.}

\section{Summary}
\label{sec:Section-5}

The offline gated correlator at GMRT can unambiguously localise pulsars within arc-second positional accuracy in a single observation. An accurate initial position facilitates follow-up with sensitive PA beam at multiple frequency bands and also accelerates the convergence of a timing model by removing covariance between long-term varying parameters, such as position and period derivative. We report an increase in the bandwidth of the offline gated correlator from 16.67 to 33.33 MHz and validate its functionality. This significant increase in correlator bandwidth also opens up new aspects, such as investigating off-pulse emission of pulsars including MSPs.

Using gated imaging and multiple PA beam search, we successfully localised J1242$-$4712 and J2101$-$4802 with positional accuracy within 1 arc-second at distances of 51$'$ and 18$'$ from the GHRSS survey pointing centres.{J1242$-$4712 and J2101$-$4802} show more than 200 and{450} detection significance, respectively, in PA beam observations of band-3 for an hour of integration. The detection S/N of J1242$-$4712 does not differ significantly from band-3 to band-4. However, more than an hour of observation is required to achieve S/N of more than{200} for J2101$-$4802 in band-4 indicating its steep spectral nature{between 300 to 750 MHz}.
In band-5, both pulsars show weaker detection significance of $\sim${40}.

J1242$-$4712 and J2101$-$4802 have the most precise ToAs in band-4 and band-3, respectively. The minimum ToA uncertainty for J1242$-$4712 is 444 ns and for J2101$-$4802 it is{539} ns. Both pulsars have the best DM precision in GMRT band-3, specifically $9\times 10^{-5}\, pc\,cm^{-3}$ (for J1242$-$4712) and{$6.3\times 10^{-5}\,pc\,cm^{-3}$} (for J2101$-$4802). We compared the ToA/DM precision of the two MSPs (in this work) to the best median ToA/DM precision of 47 NANOGrav PTA MSPs. This comparison shows that the ToA/DM precision of the two MSPs are comparable to that of the 47 PTA MSPs.{The long-term follow-up timing of these two MSPs with the PA beam in band-3 of the GMRT is currently ongoing. The results from the timing studies will be reported in the future GHRSS publications.} 

\cite{2022arXiv220104386S} recently reported the long-term wide-band timing of four GMRT discovered MSPs. Considering band-3 detection S/N, those four MSPs are roughly four times weaker than the two MSPs reported in this paper. The timing of those four MSPs resulted in $\sim\mu s$ order timing precision from 2$-$4 years of timing baseline. {The arc-second localisation of J1242$-$4712 and J2100$-$4802 will facilitate in the quicker convergence of their model parameters and the achievement of a phase-coherent solution from the follow-up observations.} We expect to get even better timing precision from the follow-up timing for the two newly localised MSPs, which will be reported in future publications. An increase in the number of well-timed MSPs in the PTA experiments may aid to the detection significance towards the  stochastic GW signals hidden in timing noise.

\section{Acknowledgments}
We gratefully acknowledge the Department of Atomic Energy, Government of India, for its assistance under project no. 12-R$\&$D-TFR-5.02-0700. The GMRT is run by the National Centre for Radio Astrophysics - Tata Institute of Fundamental Research. We thank the GMRT telescope operators for their assistance with the observations. We also thank Timothy T. Pennucci and Scott M. Ransom for providing the data files for Figure 2 of \cite{2021ApJS..252....5A} allowing us to comapre these two new MSPs with the NANOGrav samples.

\end{document}